\documentclass[%
 reprint,
 amsmath,amssymb,
 aps,
]{revtex4-2}

\usepackage{graphicx}
\usepackage{dcolumn}
\usepackage{bm}
\usepackage{hyperref}

\begin{document}

\preprint{APS/123-QED}

\title{Structural characteristics in network control of molecular multiplex networks}

\author{Cheng Yuan}
\author{Zu-Yu Qian}
\author{Shi-Ming Chen}
\author{Sen Nie}%
\email{niesen@ecjtu.edu.cn}
\affiliation{School of Electrical and Automation Engineering, East China Jiaotong University, Nanchang, Jiangxi 330013, People's Republic of China}

\date{\today}

\begin{abstract}
Numerous real-world systems can be naturally modeled as multilayer networks, enabling an efficient way to characterize those complex systems. Much evidence in the context of system biology indicated that the collections between different molecular networks can dramatically impact the global network functions. Here, we focus on the molecular multiplex networks coupled by the transcriptional regulatory network (TRN) and protein-protein interaction (PPI) network, exploring the controllability and energy requiring in these types of molecular multiplex networks. We find that the driver nodes tend to avoid essential or pathogen-related genes. Yet, imposing the external inputs to these essential or pathogen-related genes can remarkably reduce the energy cost, implying their crucial role in network control. Moreover, we find that lower minimal driver nodes as well as energy requiring are associated with disassortative coupling between TRN and PPI networks. Our findings in several species provide comprehensive understanding of genes' roles in biology and network control. 
\end{abstract}

\maketitle

\section{Introduction}
Networks are prevalent in exploring  the phenomenons and principles of our daily lives, e.g., traffic~\cite{albert2002statistical,newman2003structure}, financial~\cite{mantegna1999introduction,jiang2019multifractal}, biological~\cite{yan2017network,power2011functional,jeong2001lethality,lee2002transcriptional} and social systems~\cite{deng2018peer,wang2013evolution}. As the ultimate goal to explore these systems is to drive them into desired states, numerous advances have been achieved against network control~\cite{liu2011controllability,liu2016control,li2019controlling}. Controllability as the first step of controlling, quantifies if the system can be driven from any initial state to any desired state within finite time with finite external inputs. Liu et al.~\cite{liu2011controllability} creatively combined the structural controllability theory with complex networks and proposed a method to determine the minimal number of inputs (driver nodes) to fully control the directed networks. Exact controllability is another framework to analyze the controllability of complex networks with arbitrary structures and link weights~\cite{yuan2013exact}. Meanwhile, the minimal inputs~\cite{posfai2013effect,xiang2019advances}, optimal control strategy~\cite{gao2014target,xiao2014edge}, and structural characteristics of network in controlling have been explored~\cite{nepusz2012controlling,jia2013emergence,zhao2015intrinsic,wang2015controlling}. Though the network is completely controllable in theory, it could be difficult to control it in practice due to the huge energy requiring. Consequently, much effort has been expended on various mechanisms in determining the energy cost of network control~\cite{yan2012controlling,sun2013controllability,yan2015spectrum,nie2016effect,chen2016energy,lindmark2018minimum,wang2017physical,nie2018control,pasqualetti2014controllability}.

Functioning of many systems usually requires the coupling between different types of networks~\cite{yuan2014exact,nie2015effect,lee2015towards,battiston2014structural,zhu2019investigation}. Such as traffic systems are constituted by road network, railway network, subway network and so on~\cite{sole2016congestion}. Communication systems include different on-line and offline social networks~\cite{mucha2010community,cozzo2013contact}. Biological networks also function as the consequences of complex interactions between different molecular networks~\cite{mahajan2021internetwork,maniatis2002extensive,barabasi2004network,yeger2004network}. For instance, proteins in the protein-protein interaction network (PPI) are translated from genes in the transcriptional regulatory network (TRN)~\cite{mahajan2021internetwork}.

Recently, Mahajan et al.~\cite{mahajan2021internetwork} explored the interactions between TRN and PPI networks of different species and revealed the impact of  multiplex architectures in network robustness. They found that the functionally essential genes and proteins are situated in important parts of the multiplex networks. Here, we examine the association between genes' functional characteristics and their roles in the network control. We show that imposing external inputs to these essential or pathogen-related genes can efficiently reduce the control energy, despite the minimal driver nodes set tends to avoid them. Moreover, we find that negative correlation between TRN and PPI layer can simultaneously decrease the number of driver nodes and energy requiring.  

\section{Model}
\begin{figure}
\includegraphics[width=8cm]{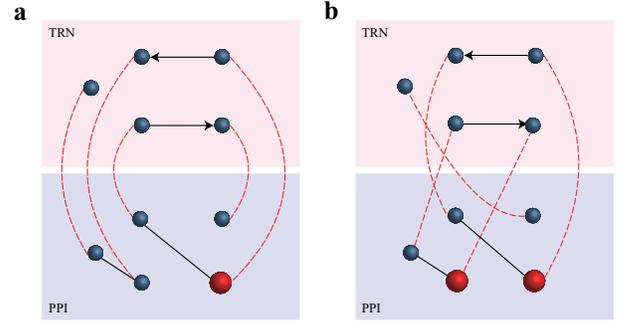}
\caption{\label{fig1} \textbf{Schematic diagram of TRN-PPI multiplex networks with different interlayer couplings}. The nodes in two layers represent the genes and proteins, respectively. The red nodes are the minimal driver nodes which are imposed on external inputs. The black directed lines represent interactions between transcription factor and its target genes in TRN layer, and the black undirected lines represent the interactions between proteins in PPI layer. The red dash lines represent the one-to-one interactions between genes and their corresponding proteins. The interlayer coupling are different in \textbf{(a)} and \textbf{(b)}, thus the minimal driver nodes set are changed.}
\end{figure}

Here, transcriptional regulatory networks (TRN) represent the interactions among transcription factors and their target genes. Edges in the TRN encode direct interactions between a transcription factor and its target genes. In protein-protein interaction networks (PPI), each node represents a protein and undirected link denotes the physical or binding interactions. Coupling links between layers represent the interaction between the genes in TRN and the proteins in PPI, in which the proteins translated from genes could also regulate other genes. The coupling between TRN and PPI layers is one-to-one correspondence, i.e., a gene in TRN is collected to a corresponding protein in PPI layer~\cite{mahajan2021internetwork}. The Schematic diagram of the multiplex networks is shown in Figure~\ref{fig1}. We employ the datasets in Ref.~\cite{mahajan2021internetwork}, which are collected from 7 species:~\emph{H. pylori}~\cite{danielli2010built,hauser2014second},~\emph{M. tuberculosis}~\cite{sanz2011transcriptional,wang2010global},~\emph{E. coli}~\cite{salgado2018using,rajagopala2014binary,das2012hint},~\emph{C. elegans}~\cite{das2012hint,chatr2017biogrid,fuxman2016gene},~\emph{A. thaliana}~\cite{das2012hint,chatr2017biogrid,jin2015arabidopsis},~\emph{M. musculus}~\cite{das2012hint,chatr2017biogrid,han2018trrust}, and~\emph{H. sapiens}~\cite{das2012hint,chatr2017biogrid,han2018trrust}. Since TRN and PPI networks have different coverage of the genome and proteome, only the genes and proteins presented in both layers are considered in our analysis. 

Formally, the dynamics of each node in TRN-PPI multiplex networks is described by
\begin{equation} 
\label{eq1}
\dot{\boldsymbol{z}}(t)=\boldsymbol{f}(\boldsymbol{z},\boldsymbol{v},t)
\end{equation}
where $\boldsymbol{z}(t)=(z_{1}(t),z_{2}(t),\dotsc,z_{N_{1}+N_{2}}(t))^T$ is the state of $N_{1}+N_{2}$ nodes at time $t$. $N_{1}$ and $N_{2}$ are the size of TRN and PPI networks, respectively. $\boldsymbol{f}(*)=[f_{1}(*),f_{2}(*),\dotsc,f_{ N_{1}+N_{2}}(*)]^T$  captures the nonlinear dynamics of each node, and $\boldsymbol{v}(t)=[v_{1}(t),v_{2}(t),\dotsc,v_{M}(t)]^T$ is the external inputs imposed on the multiplex networks. For simplicity, assuming the system is at a fixed point $\boldsymbol{z}^*$, where $\boldsymbol{f}(\boldsymbol{z}^*, \boldsymbol{v}^*,t)=0$, and $\boldsymbol{x}(t)=\boldsymbol{z}(t)-\boldsymbol{z}^*$,  $\boldsymbol{u}(t)=\boldsymbol{v}(t)-\boldsymbol{v}^*$. Equation~\ref{eq1} can be linearized as
\begin{equation} 
\label{eq2}
    \dot{\mathbf{x}}(t)={
\left[ \begin{array}{cc}
A_{11} & A_{12} \\
A_{21} & A_{22} 
\end{array} 
\right ]} \mathbf{x}(t)+B\mathbf{u}(t)
\end{equation}
where $\left[\begin{array}{cc}A_{11} & A_{12} \\A_{21} & A_{22} \end{array} \right ]\equiv{{\frac{\partial \boldsymbol{f}}{\partial \boldsymbol{z}}}\bigg|_{\boldsymbol{z}^*,\boldsymbol{v}^*}}$ is the global adjacency matrix of the network and captures the interactions between $N_1+ N_{2}$ nodes. Whereas, $A_{11}$ ($A_{22}$) describes the connections between nodes within TRN (PPI) networks and $A_{12}$ ($A_{21}$) captures the intra-connections between TRN and PPI networks. $B\equiv{{\frac{\partial \boldsymbol{f}}{\partial \boldsymbol{v}}}\bigg|_{\boldsymbol{z}^*,\boldsymbol{v}^*}}$ represents how the external inputs are imposed on nodes. 

Note that if the linear system~\ref{eq2} is locally controllable along a trajectory in state space, then the corresponding nonlinear system~\ref{eq1} is also controllable along the same trajectory. In addition, the linear control predictions are consistent with control of nonlinear dynamics~\cite{yan2017network}. Hereafter, to apply the controllability and control energy framework, we focus on the linear system~\ref{eq2}. 

\section{Methods}
\subsection{Minimal driver nodes and control energy}
A system is controllable if it can be driven from any initial state to any desired state with finite external inputs within finite time. The nodes imposed on external inputs are driver nodes, and the minimal number of driver nodes measures the controllability of a network, defined as $N_\text{D}$, and $n_\text{D}=\frac{N_\text{D}}{N}$. According to exact controllability framework~\cite{yuan2013exact}, $N_\text{D}$ can be calculated as
\begin{equation} 
\label{eq3}
N_\text{D}=\max_{i}\left(\mu\left(\lambda_{i}\right)\right)
\end{equation}
where $\lambda_{i}\left(i=1,2,\dotsc,N\right)$ represent the eigenvalues of adjacency matrix, and $\mu\left(\lambda_{i}\right)$ is the geometric multiplicity. Moreover, the minimal driver nodes set can be identified in $B$ to satisfy the equation
\begin{equation} 
\label{eq4}
rank[A-\lambda^MI_\text{N},B]=N
\end{equation}
where $\lambda^{M}$ refers to the eigenvalue according to the maximum geometric multiplicity $\mu(\lambda^{M})$. It was noted that the rank of matrix $[A-\lambda^MI_\text{N},B]$ is determined by the number of linearly independent rows. By performing the elementary column transformation on the matrix $A-\lambda^MI_\text{N}$, we can obtain the linearly dependent rows. Therefore, the external inputs described by $B$ should be imposed on the rows to eliminate all linear correlations to make the matrix $[A-\lambda^MI_\text{N},B]$ full-ranked, and the corresponding minimal driver nodes set can be identified~\cite{yuan2013exact}.

On the basis of optimal control theory~\cite{rugh1996linear}, the energy required to control the system is $E(t_0,t_f)=\int_{t_0}^{t_f}\|\mathbf{u}(t)\|^2\mathrm{d}t$. If the initial state $\mathbf{x}_0=0$ at $t=0$, the minimal control energy is 
\begin{equation} 
\label{eq5}
E(t_f)=\mathbf{x}_{t_f}^TW^{-1}(0,t_f)\mathbf{x}_{t_f}
\end{equation}
where $W(0,t_f)=\int_{0}^{t_f}e^{At}BB^Te^{A^Tt}\mathrm{d}t$ is the Gramian matrix. As the control energy decays to a nonzero stationary value with increasing time $t$, the control energy discussed here is $E (t_f\rightarrow\infty)$, and the Gramian matrix is $W (0,t_f\rightarrow\infty)$. We set the elements as $A_{ii}=-(\delta + \sum_{j=1}^{N}A_{ij})$, where $\delta=0.25$ to ensure the stability of the whole system.
 
\subsection{Degree-degree correlation}
Associativity between TRN and PPI layers is measured by the Pearson’s correlation coefficient $\rho$~\cite{foster2010edge}
\begin{equation} 
\label{eq6}
\begin{aligned}
\rho &=cor(k_\text{out},K)\\
     &={{\sum_{i=0}^{n} (k_{\text{out}}(i)-\overline{k}_\text{out})\sum_{i=0}^n(K(i)-\overline{K})}\over{\sqrt{\sum_{i=0}^n(k_{\text{out}}(i)-\overline{k}_\text{out})^2}\sqrt{\sum_{i=0}^n(K(i)-\overline{K})^2}}}
     \end{aligned}
\end{equation}
where $k_\text{out}$ is the out-degree of node in the TRN network, $K$ is the degree of node in the PPI network and $n$ is the number of nodes in each layer.

\subsection{Simulated annealing}
To finely tune the degree-degree correlation coefficient $\rho$ of multiplex networks, we adopt the simulated annealing algorithm~\cite{kirkpatrick1983optimization}:

1. Randomly shuffle gene labels in the TRN network, while keep protein labels in the PPI network unchanged. Then calculate the absolute difference between the shuffled and desired degree-degree correlations as $\Delta^{P}=\vert cor_{k,K}^{P}-cor_{k,K}^{D} \vert$, where $cor_{k,K}^{P}$ is the present correlation between nodes with $k_\text{out}$ degree in TRN and nodes with $K$ degree in the PPI. $cor_{k,K}^{D}$ is the desired correlation. Save the present gene labeling in the TRN network.

2. Randomly select a subset of size $N$ (we use $N = 10$) from the genes in the TRN network, and shuffle their labels randomly. Save this modified labeling of the TRN as a new sampling. Then define the new difference as $\Delta^{*}=\vert cor_{k,K}^{*}-cor_{k,K}^{D} \vert$, where $cor_{k,K}^{*}$ is degree-degree correlation of the present network after randomly shuffling $10$ genes labels, and $cor_{k,K}^{D}$ is the desired correlation.

3. Calculate the difference $\Delta=\Delta^{P}-\Delta^{*}$. 
If $\Delta \ge 0 $, accept the labeling proposed in step 2 as the current labeling for TRN. Otherwise, accept the labeling proposed in step 2 with probability $e^{\Delta \over T}$, where $T=T_0e^{-\lambda L}$ is temperature, $L$ is the iteration number and $\lambda$ is rate parameter. The parameters are set as $T_0=1,000$, $\lambda=0.01$.

4. If $\Delta^{P}$ is smaller than the defined value (here we set the defined value as $0.01$) then stop, else repeat from the step 2.

\section{Results}

\begin{figure*}
\includegraphics[width=18cm,height=3cm]{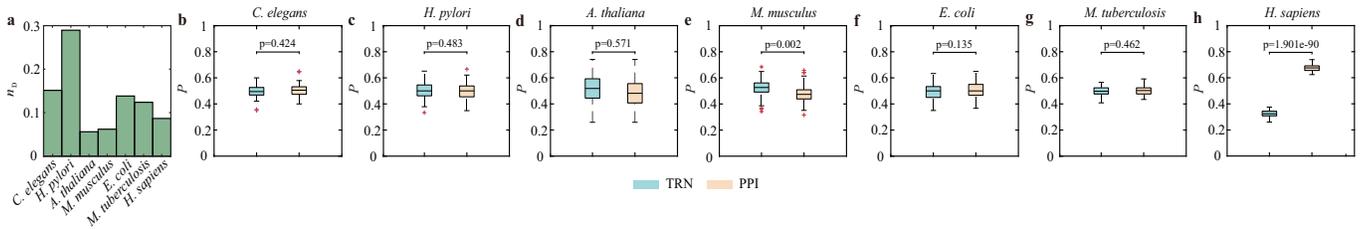}
\caption{\label{fig2} \textbf{Controllability and driver nodes distribution of TRN-PPI multiplex networks.} \textbf{(a)} The fraction of minimal driver nodes of seven species. \textbf{(b)-(h)} are the fractions of minimal driver nodes in two layers for \textbf{(b)} \emph{C. elegans}, \textbf{(c)} \emph{H. pylori}, \textbf{(d)} \emph{A. thaliana}, \textbf{(e)} \emph{M. musculus}, \textbf{(f)} \emph{E. coli}, \textbf{(g)} \emph{M. tuberculosis}, and \textbf{(h)} \emph{H. sapiens}. We obtain 100 minimal driver nodes sets, and calculate the average value of the driver nodes fractions in two layers. P-value is calculated by t-test.}
\end{figure*}

To reflect the controllability of these TRN-PPI multilayer molecular networks, we firstly examine the minimal driver nodes $N_\text{D}$ to achieve fully control of the network in each species (Figure~\ref{fig2}(a)). The characteristics of networks for each species is shown in Table~\ref{tab1}. We find that overall those networks display lower $n_\text{D}\ (\sim0.15)$, indicating that we need to independently control about $15\%$ nodes to fully control them. With the smallest average degree, \emph{H. pylori} yields the highest $n_\text{D}$. This is consistent with previous finding that sparse network is more difficult to control. Then, we examine the distribution of driver nodes between two layers, implying that these driver nodes do not have any particular preference between two layers for most of species (Figure~\ref{fig2}(b)-(h)). However, we find that the fraction of driver nodes associated with PPI layer is significantly larger than that of TRN layer for \emph{H. sapiens} (p-value$<$0.001). To understand such difference, we calculate the average degree $\langle k \rangle$ of TRN and PPI layer, respectively (Table~\ref{tab1}), showing that the difference of average degree between two layers is most obviously in \emph{H. sapiens} species. The density difference between two layers leads to the uneven distribution of driver nodes in TRN and PPI networks.
   
\begin{table}[h]
\caption{\label{tab1} \textbf{The characteristics of the TRN-PPI multiplex networks for seven species.}}
\begin{ruledtabular}
\begin{tabular}{lrrrrr}
Species  & $N$ & $L_\text{TRN}$ & $L_\text{PPI}$ & $\langle k \rangle_\text{TRN}$ & $\langle k \rangle_\text{PPI}$\\
\hline
\it{C. elegans} & 628 & 821 & 285 & 2.62& 1.82\\
\it{H. pylori}  & 228 & 156 & 76 & 1.37 & 1.33\\
\it{A. thaliana} & 482 & 385 & 398 & 1.60 & 3.30\\
\it{M. musculus} & 920 & 1,269 & 518 & 2.76 & 2.25\\
\it{E. coli} & 434 & 147 & 154 & 0.68 & 1.42\\
\it{M. tuberculosis} & 1,542 & 958 & 946 & 1.24 & 2.45\\
\it{H. sapiens} & 3,172 & 4,838 & 4,280 & 3.05 & 5.40
\end{tabular}
\end{ruledtabular}
\end{table}

\begin{figure}
\includegraphics[width=8cm]{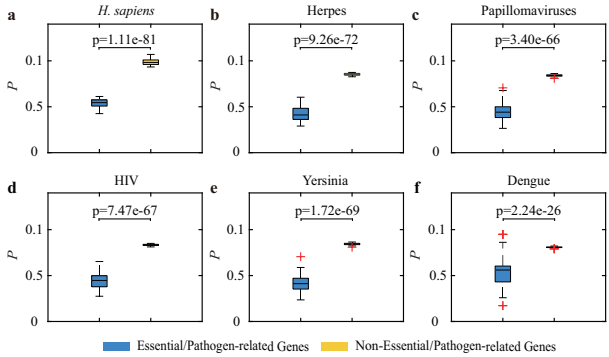}
\caption{\label{fig3} \textbf{Distribution of driver nodes in different gene categories.} \textbf{(a)} \emph{H. sapiens}, \textbf{(b)} Herpes, \textbf{(c)} Papillomaviruses, \textbf{(d)} HIV, \textbf{(e)} Yersinia, \textbf{(f)} Dengue. We obtain 100 minimal driver nodes sets, and calculate the fraction of essential or pathogen-related genes from driver nodes set in all essential or pathogen-related genes (left, blue box), and non-essential or nonpathogen-related genes from driver nodes set in all non-essential or nonpathogen-related genes (right, yellow box), respectively, then get the average value. P-value is calculated by t-test.}
\end{figure}

The essential and pathogen-related genes are critical for survival and healthy status of an organism, thus conveying particular topological characteristics in the corresponding multiplex networks~\cite{mahajan2021internetwork}. Hence, we examine whether those essential or pathogen-related genes are associated with prominent role in the network control. To achieve this, 
we adopt the gene categories in Ref.~\cite{mahajan2021internetwork}, where   
essential genes for human species are collected from the Online GEne Essentiality (OGEE) database~\cite{chen2016ogee,chen2012ogee}, and five human pathogens from a publicly available database HPIDB 3.0~\cite{kumar2010hpidb,ammari2016hpidb}.
Then, we compare the fractions of essential or pathogen-related genes and non-essential or nonpathogen-related genes are selected as the driver nodes, respectively. Figure~\ref{fig3} indicates that proportion of essential or pathogen-related genes selected as driver nodes is significantly lower than that of non-essential or nonpathogen-related genes (p-value$<$0.001), regardless of species. The fundamental role of these essential or pathogen-related genes is usually reflected as the higher degree in TRN layer. Yet, most of the driver nodes tend to be the lower degree ones in control~\cite{liu2011controllability}.  

\begin{figure}
\includegraphics[width=8cm]{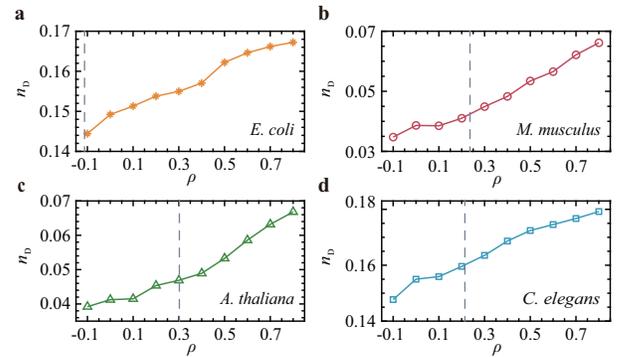}
\caption{\label{fig4}\textbf{The minimal number of driver nodes as functions of degree-degree correlation of interlayers for TRN-PPI multiplex networks.} \textbf{(a)} \emph{E. coli}, \textbf{(b)} \emph{M. musculus}, \textbf{(c)} \emph{A. thaliana}, \textbf{(d)} \emph{C. elegans}. The gray dash lines show the real value of interlayers degree-degree correlation. Each data point is the mean of 100 independent realizations.}
\end{figure}

\begin{figure}
\includegraphics[width=8cm]{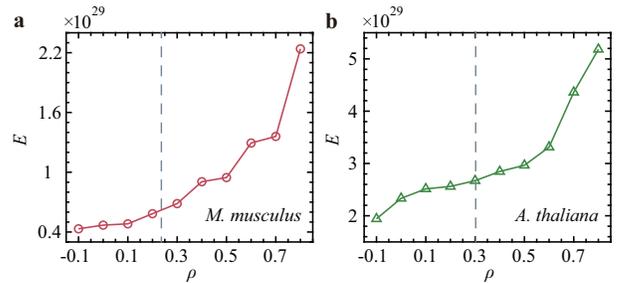}
\caption{\label{fig5} \textbf{Control energy as functions of degree-degree correlation of interlayers for TRN-PPI multiplex networks.} \textbf{(a)} \emph{M. musculus}, \textbf{(b)} \emph{A. thaliana}. The initial state is $x_0=[0,0,\dotsc,0]^T$ and final state is $x_{t_f}=[1,1,\dotsc,1]^T$. The number of driver nodes for two eukaryotes are $570$ and $200$, respectively. The gray dash lines show the real value of interlayers degree-degree correlation. Each data point is the mean of 100 independent realizations.}
\end{figure}

We further explore the impact of interlayer correlation on the controllability of networks. Note that the interactions between TRN and PPI networks have already been established by their biological relationships. The purpose of our analysis is to reveal whether the underlying coupling pattern in reality can be partially explained in a network control manner. Therefore, we finely tune the correlation between TRN and PPI layer through simulated annealing algorithm, finding that more driver nodes will be required as the network is transited from disassortative to assortative (see Figure~\ref{fig4}). We note that the real interlayer correlation for the four species are mostly positive.
We also assess how degree correlation determines the energy requiring in network control. We again finely tune the interlayer correlation by rewiring the links. To eliminate the impact of different driver nodes sets, all nodes are independently controlled. We find that control energy in terms of degree correlation displays the similar trajectory as we observed in $n_\text{D}$ (in Figure~\ref{fig4}), i.e., multiplex networks with disassortative coupling between layers are easier to control than that of assortative coupling (see Figure~\ref{fig5}).  

Finally, we examine whether those essential genes can provide critical role in the energy requiring for TRN-PPI multiplex network control. Subsequently, we propose three driver nodes selection strategies with given $N_\text{D}$: (1) All essential or pathogen-related genes, saying $N_\text{E}$ are selected as driver nodes, then the remaining $N_\text{D}-N_\text{E}$ driver nodes are selected randomly. (2) $N_\text{D}$ driver nodes are selected based on descending order of nodes degree. (3) All $N_\text{D}$ driver nodes are select at random. Interestingly, we find that the control energies of these three strategies exhibit quite consistent implication that controlling those essential or pathogen-related genes yields the lowest energy requiring. These results imply the critical role of essential or pathogen-related genes in the network control, despite the driver nodes tend to avoid them.

\begin{figure}
\includegraphics[width=8.5cm,height=5.3cm]{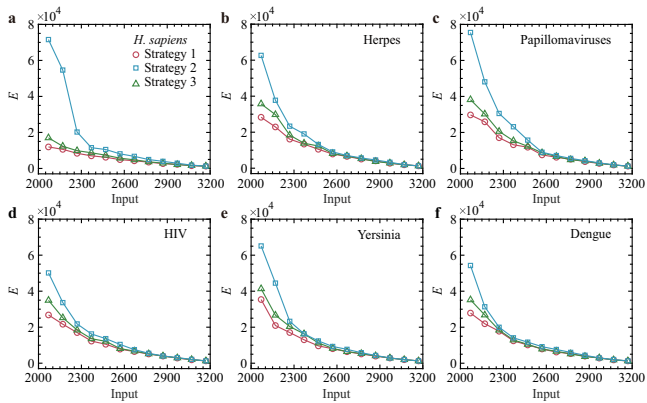}
\caption{\label{fig6} \textbf{Control energy as functions of the number of driver nodes by three strategies.} \textbf{(a)} \emph{H. sapiens}. We choose all the $1,340$ essential genes as driver nodes for \emph{H. sapiens}, and all the pathogen-related genes as driver nodes for five different pathogens in \textbf{(b)-(f)}. The number of pathogen-related genes is $414$ for \textbf{(b)}, $340$ for \textbf{(c)}, $291$ for \textbf{(d)}, $340$ for \textbf{(e)}, and $116$ for \textbf{(f)}.}
\end{figure}

\section{conclusion}
Interactions between networks are ubiquitous in molecular networks. Here, we focus on multiplex networks coupled with transcriptional regulatory network (TRN) and protein-protein interaction (PPI) network, to explore the minimal driver nodes and the energy requiring for fully control. The results indicate there is no obvious preference for driver nodes distribution in two layers, and the driver nodes are more likely to avoid the essential or pathogen-related genes. Besides, the TRN-PPI networks with positive degree-degree correlation of interlayer need more driver nodes and more energy to achieve fully control. By comparing different strategies of selecting driver nodes, we find the TRN-PPI networks requiring the lowest energy to reach the desired state by driving essential or pathogen-related genes. Our work bridge the gap between structural characteristics of molecular multiplex networks and the network control. It will be helpful for understanding the essential genes' function in biology and network control.

\begin{acknowledgments}
This work was supported by the National Natural Science Foundation of China (No.61763013), The Natural Science Foundation of Jiangxi Province (No.20202BABL212008), and the Jiangxi Provincial Postdoctoral Preferred Project of China (No.2017KY37). 
\end{acknowledgments}

\bibliography{apssamp}

\end{document}